\documentclass{article}
\usepackage{spconf,amsmath,graphicx,hyperref}
\usepackage{cite}
\usepackage{amssymb,amsfonts}
\usepackage{algorithmic}
\usepackage{algorithm}
\usepackage{textcomp}
\usepackage{xcolor}
\usepackage{subcaption}
\usepackage{booktabs}
\title{Federated Coreset Subsampling for Distributed Memory Banks}
\name{Jason Witzemann$^{1}$, Daniel Pr\"oll$^{1,3}$, Sebastian Hegenbart$^{1,2}$}
\address{$^{1}$Digital Factory Vorarlberg GmbH, Dornbirn, Austria\\
         $^{2}$Vorarlberg University of Applied Sciences, Dornbirn, Austria\\
         $^{3}$University of Salzburg, Salzburg, Austria }

\begin{document}

\maketitle

\begin{abstract}
%Federated learning is used to train distributed models. In anomaly detection settings few-show capable methods rely on pre-trained feature extractors skipping the classical training step entirely. Instead, such methods construct feature-agnostic coresets to model the nominal data class. 

In real-world scenarios, industrial processes can span multiple factories and goods are often produced at multiple sites, leading to a shared yet non-identical distribution of data across locations. To address this, we propose two federated approaches for i) combining existing coresets (\textit{Double Subsampling}) and ii) creating new coresets (\textit{Federated Subsampling}) in a distributed anomaly detection setting. In particular, we adapt the memory bank based PatchCore algorithm to a federated setting, resulting in increased cold-start performance, resilience to domain- and sensor-shift, and resilience to unreliable actor participation, a common problem in classical, network based federated learning strategies. 
\end{abstract}


\section{Introduction}
Distributed and small-lot production settings (e.g., molds production, aerospace components, contract electronics) generally require few-shot capable methods for anomaly detection due to the limited amount of available training samples as well as the cost and effort of data distribution between sites. Small-lot production suffers from poor cold-start performance, while in distributed settings a consistent detection model is a requirement at all sites to ensure product quality. While zero-shot models are gaining interest in anomaly detection and could potentially be used in such settings, they currently lack the depth and accuracy of detection systems utilizing a smaller but non-zero corpus of nominal training images. Combining data from federated production facilities offers access to a richer training corpus, potentially increasing model accuracy. This motivates the contributions of this paper as follows: We extend PatchCore \cite{roth_towards_2022} to harness the strengths of few-shot capable memory bank based approaches in a federated setting by introducing two distinct approaches. We introduce i) \textit{Double Subsampling}, a hierarchical approach to combining federated memory banks, which can utilize multiple existing local models to create a global anomaly detection model, and ii) \textit{Federated Subsampling}, a federated approach for creating memory banks. Both approaches leverage the few-shot capability of PatchCore, improving  cold-start performance, implicitly correcting for domain- and sensor-shift over multiple production sites.

\section{Related Work}
\textbf{Federated Learning} leverages data across multiple actors to train a shared model. As each actor's local data distribution can differ from the aggregated global distribution, averaging locally trained model updates does not guarantee convergence towards the same optimum as centralized training \cite{wang_tackling_2020, li_federated_2020, mcmahan_communication-efficient_2017}. This non-IID setting is particularly important in an industrial anomaly detection context.

The memory bank paradigm is inherently well suited to sidestep this aggregation problem, since it does not rely on gradient-based training and instead directly uses features from pre-trained extractors based on nominal training samples. Roth et al. \cite{roth_towards_2022} proposed PatchCore, a state-of-the-art memory bank approach for practical industrial anomaly detection. The strengths of this approach are the few-shot possibility, the small inference model size, and the ability for rapid re-training. While other approaches \cite{zavrtanik_draem_2021, yu_fastflow_2021, jeong_winclip_2023, batzner_efficientad_2024} show competitive results, they utilize neural networks for inference, need extensive training times, or rely on a rich corpus of nominal training samples to achieve competitive performance. Chen et al. \cite{chen_feddymem_2025} first proposed the concept of employing federated memory banks (FedDyMem). A global memory bank for inference is constructed by training a local network to project local features towards a global distribution. While FedDyMem combines the memory bank for easy inference with known federation strategies, the approach still utilizes training of a neural network as the core component for creating the memory bank, leaving room for longer training runs and non-IID data drifts. To close this gap, we propose \textit{Federated Subsampling} and \textit{Double Subsampling} for constructing federated memory banks based on few-shot capable methods, enabling multiple actors to jointly construct a shared coreset of nominal features under the few-shot cold-start constraint. In the following, we detail the architectures and creation processes.

\section{Preliminary}
\label{sec:preliminary}
We revisit the preliminary context for introducing our approaches. This includes the federated learning setup, which holds for both proposed methods, \textit{Double Subsampling} and \textit{Federated Subsampling}, as well as the explanation of PatchCore \cite{roth_towards_2022}.

\textbf{Federated Setting}
In the proposed federated learning setting, there exist $N$ actors $\{C_1, C_2, \dots, C_N\}$ and a single Server $S$. Each actor $n \in N$ has a local set of nominal RGB images, contained in a dataset $I^n = \{I^n_1, I^n_2, \dots\}$, where $I^n_i \in \mathbb{R}^{224 \times 224 \times 3}$. We assume that the images across all actors share the category of product but vary in environmental settings such as background, lighting, scale, or used sensors.  

\textbf{Coreset Subsampling}
is used to identify and collect the most informative feature vectors to create a memory bank and is a core component of PatchCore. The algorithm is feature-agnostic and only assumes the existence of a meaningful metric between feature vectors. 
Let $\mathcal{M}$ denote the set of all features extracted from a corpus of nominal training data. Coreset subsampling reduces $\mathcal{M}$ to a coreset $\mathcal{M}_C$ with $\mathcal{M}_C \subset \mathcal{M}$ to
represent the original set as closely as possible with the least amount of features necessary. This implicitly acts as regularization and improves inference performance. The cardinality of the coreset is a hyper-parameter and depends on the specific use-case. 

The reduction process is based on a greedy iterative approximation of a \textit{minimax facility location} coreset selection \cite{sinha_small-gan_2020, sener_active_2018}.
Initially, a random set $\mathcal{T}$ of $k$ features  $k \ll |\mathcal{M}|$ is sampled. For each element in $\mathcal{M}$, the nearest neighbor in $\mathcal{T}$ is identified (L2 is used in PatchCore). 
Next, the element $x\in \mathcal{M}$ with the largest distance to its nearest neighbor in $\mathcal{T}$ is added to $\mathcal{M}_C$.
Once $\mathcal{M}_C \not= \emptyset$, the vector in $\mathcal{M}$ with the largest distance to its nearest neighbor in $\mathcal{M}_C$ is added to $\mathcal{M}_C$ until the coreset target size is reached.

\section{Methods}
 The proposed methods in this work, \textit{Double Subsampling} and \textit{Federated Subsampling}, depart from the traditional federated parametric training regime. Instead, they leverage the non-parametric memory bank paradigm.

\textbf{Double Subsampling}
In contrast to a traditional iterative approach, the proposed \textit{Double Subsampling} method uses a hierarchical coreset subsampling process to combine existing coresets. We assume that each actor owns a local coreset $\mathcal{M}_{C,i}$ (see~\autoref{sec:preliminary}). In practice, this is a reasonable assumption, as the feature extraction and inference steps are usually  performed actor-side.  For each actor, the local coreset $M_{C,i}$ is propagated to the server and aggregated into a single global memory bank $\mathcal{M}_S = \bigcup \mathcal{M}_{C,i}$. 
Each $M_{C,i}$ contains a distribution of the most informative and representative local features. The resulting union over all coresets $\mathcal{M}_S$ can be seen as a global representation of the entire distributed data corpus. 
As a regularization step, and to improve computational performance, we perform coreset subsampling on $\mathcal{M}_S$ at the server again, resulting in a final global coreset $M^*$, which is propagated to all actors $C_i$.

% This table is here for formatting reasons only, it belongs to the results
\begin{table*}[t!]
    \centering
    \caption{Comparison of Pixel-level AUROC of PatchCore, Double Subsampling, and Federated Subsampling across MVTec-AD categories}
    \label{tab:mvtec_results}
    \resizebox{\textwidth}{!}{
    \begin{tabular}{l c ccccccccccccccc}
    \toprule
    \textbf{Method} & \textbf{Total} & Bottle & Cable & Capsule & Carpet & Grid & Hazelnut & Leather & Metal Nut & Pill & Screw & Tile & Toothbrush & Transistor & Wood & Zipper \\
    \midrule
    PatchCore             & \textbf{95.9} & 98.1 & 96.0 & 98.2 & 98.7& 95.2& 97.8& 99.1& 94.8& 98.3& 90.2& 94.6& 98.0& 89.9& 92.8& 97.5 \\
    \midrule
    Federated Subsampling & \textbf{95.9} & 98.1 & 96.0& 98.2& 98.7& 95.2& 97.8& 99.1& 94.8& 98.3& 90.2& 94.6& 98.0& 89.9& 92.8& 97.5\\
    Double Subsampling    & \textbf{95.8} & 98.0 & 95.8& 97.9& 98.6& 95.3& 97.7& 99.1& 94.7& 98.0& 88.1& 94.8& 98.0& 90.6& 92.7& 97.5 \\
    \bottomrule
    \end{tabular}
    }
\end{table*}

\textbf{Federated Subsampling}
\label{sec:FedSub}
More in line with existing federated learning strategies, the proposed \textit{Federated Subsampling} employs an iterative approach, where in each iteration, all actors calculate a potential candidate for the best vector to add to the global coreset $\mathcal{M}^*$. The number of iterations performed in this approach equals the final number of vectors in the global memory bank, which is a problem-specific figure that can range from a percentage of the total amount of features to a fixed value. As a greedy iterative approximation is used to select the next feature for the memory bank, we utilize the iterative properties of the coreset reduction process and established federated learning frameworks to construct a strategy to add features iteratively to a global coreset $\mathcal{M}^*$. The candidate proposal for each actor works similarly to the coreset reduction process, where each actor proposes the vector with the largest distance to its nearest neighbor in $\mathcal{M}_{C,i}$. The proposed vectors $v_i$ (see \autoref{eq:candidate}) and the corresponding distances $d_i=min_{y \in \mathcal{M}_{C,i}}\|v_i-y\|_2$ to the nearest neighbors in $\mathcal{M}_{C,i}$ are aggregated on the server per actor as $V = \{(v_1, d_1),\dots,(v_i, d_i)\}$.

 On the server, the distances in each tuple are used to select the winning candidate, following the last step of the coreset reduction process. The feature part of the tuple is used for the coreset and is the object that is added to the global coreset $\mathcal{M}^*$ as well as propagated to each actor's coreset $\mathcal{M}_{C,i}$. This candidate-proposal and -selection process is repeated until no more feature vectors need to be added to the memory bank. As this approach is a distributed version of the greedy coreset subsampling process of PatchCore, we show (\autoref{experiments}) that the resulting $M^*$ is equal to a locally created memory bank, created by the PatchCore algorithm. To improve communication efficiency, this method can be extended to propose $\lambda$  candidates per proposal instead of one. For this, each actor $C_i$ locally iteratively identifies a sequence of candidates $\{v_{i,1}, v_{i,2}, \dots, v_{i,\lambda}\}$. 

After each update, every actor is presented with an additional entry to the global coreset. A batch-based approach is used as a tradeoff between the required communication rounds and the increased deviation to the optimal coreset by \textit{temporarily updating} the local coresets after selecting $v_{i,j}$ using the standard greedy approach.
Each actor pretends that its candidate $v_{i,j}$ has won the selection round and calculates the successor $v_{i,j+1}$ based on this assumption.
 \begin{align}
    \label{eq:candidate}
    v_i = \operatorname*{arg\,max}_{x \in \mathcal{M}_i} \min_{y \in \mathcal{M}_{C,i}} \|x-y\|_2
\end{align}

 After a batch of candidates $\{v_{i,1}, v_{i,2}, \dots, v_{i,\lambda}\}$ is selected locally, a tuple with the aggregated distance $D_i = \sum_{j=1}^{\lambda}d_{i,j}$ is sent to the server as $V_i = (\{v_{i,1}, v_{i,2}, \dots, v_{i,\lambda}\}, D_i)$. From there, the original approach of greedily selecting features based on the largest total distance is applied, adding all features of the winning batch to the global coreset and propagating them to the actors. 

A strength of the proposed methodology is its compatibility to the standard non-federated PatchCore method and memory bank based methods in general. Both proposed approaches act merely on the coreset subsampling part, effectively enabling the use of the standard inference and feature-extraction process. In particular, we  follow the proposed process using neighbor-aware anomaly score calculation per patch with bi-linear upscaling to pixel-level and a gaussian blur for smoothing. 
Inference is performed actor-side on the coresets propagated from the server but can theoretically be performed server-side, propagating only the anomaly-score back to the actor. Server-side inference could prove useful in scenarios with untrusted or potentially malicious actors contributing to a shared memory bank.

\section{Experiments}
\label{experiments}

 All experiments are conducted using the standard benchmark dataset for structural anomalies, MVTec AD \cite{bergmann_mvtec_2019}. This dataset is comprised of 15 categories, covering 5,354 high-resolution images, with 73 different defect types. The methods and experiments were implemented using \textit{PyTorch}. \textit{ray} was used for simulating multiple actors. The feature-extractor backbone used in the experiments is a WideResNet50 using \textit{layer-2} and \textit{layer-3} with adaptive average pooling, resulting in a 1000-dimensional locally aware patch feature. Following the idea of \cite{sinha_small-gan_2020} and \cite{dasgupta_elementary_2003}, we performed the coreset selection process as explained in \autoref{sec:preliminary} using dimensionality-reduced feature vectors of 100 dimensions. This dimensionality reduction was performed with a randomly initialized linear projection.
 
\textbf{AUROC compared to Baseline PatchCore}
As our two proposed methods, \textit{Federated Subsampling} and \textit{Double Subsampling}, are extensions of the PatchCore algorithm, we compared the pixel- and image-level Area Under the Receiver Operating Characteristic (AUROC) score to validate the accuracy of the distributed extension compared to the original localized algorithm. As seen in \autoref{tab:mvtec_results}, constructing coresets from scratch using the \textit{Federated Subsampling} approach results in equal scores across all categories compared to the original algorithm, indicating equally constructed coresets between our approach and the baseline. However, the \textit{Double Subsampling} approach yields similar, yet different performance, deviating slightly from the baseline depending on the category. This is likely an effect caused by the local subsampling step, rejecting local candidates that could improve the global memory bank and vice versa. We consequently focus the follow-up experiments on the \textit{Federated Subsampling}. 

\begin{figure}[b!]
    \centering
    \includegraphics[width=0.8\columnwidth]{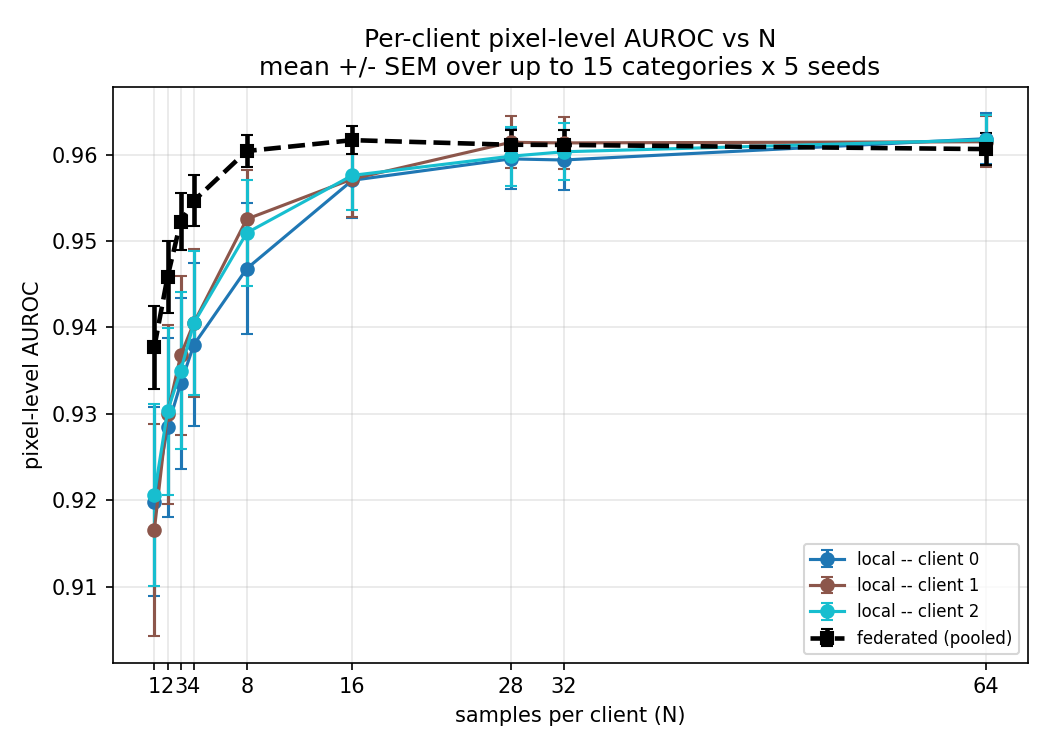}
    \caption[Sample Improvement]{Federated Subsampling and per-actor pixel-level AUROC comparison with increasing training sample size}
    \label{fig:sample_improv}
\end{figure}

 \textbf{Cold-start performance}
 We setup a cold-start scenario to illustrate the performance increase gained through \textit{Federated Subsampling} by participating as an actor towards constructing a shared coreset. Leveraging the larger available training corpus AUROC scores are increased, as shown in \autoref{fig:sample_improv}. The fewer samples that are available locally, the more the data federation benefits the quality of the global coreset. While in a single-sample training corpus per actor the image-level AUROC improves from $84.2$ to $89.2$, the improvement with 64 available samples per actor is marginal, going from $97.1$ to $97.4$. This is an expected behavior as the quality of the global- and local-memory banks converge with larger training corpi. The results suggest that the \textit{Federated Subsampling} approach is beneficial for cold-start problems, such as pilot series production or small-lot production.

\textbf{Non-IID adaptation}
\textit{Federated Subsampling} can be useful when dealing with domain- and sensor-shifts in data. Assuming a distribution over domains and sensors exists across actors, the created global memory banks are robust against such effects. 
This is in contrast to traditional localized memory bank based methods, which require re-buildling a stale memory bank with the new data distribution to avoid degradation in accuracy. 

\begin{figure}[ht!]
    \centering
    \includegraphics[width=0.85\columnwidth]{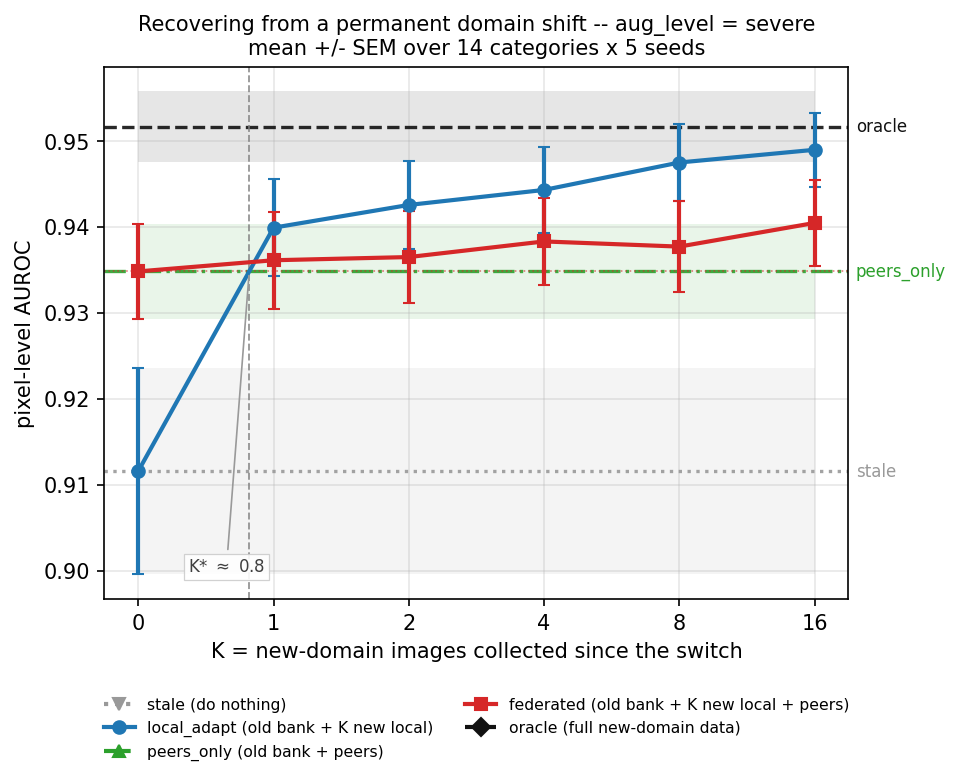}
    \\[0.5em]
    \includegraphics[width=0.85\columnwidth]{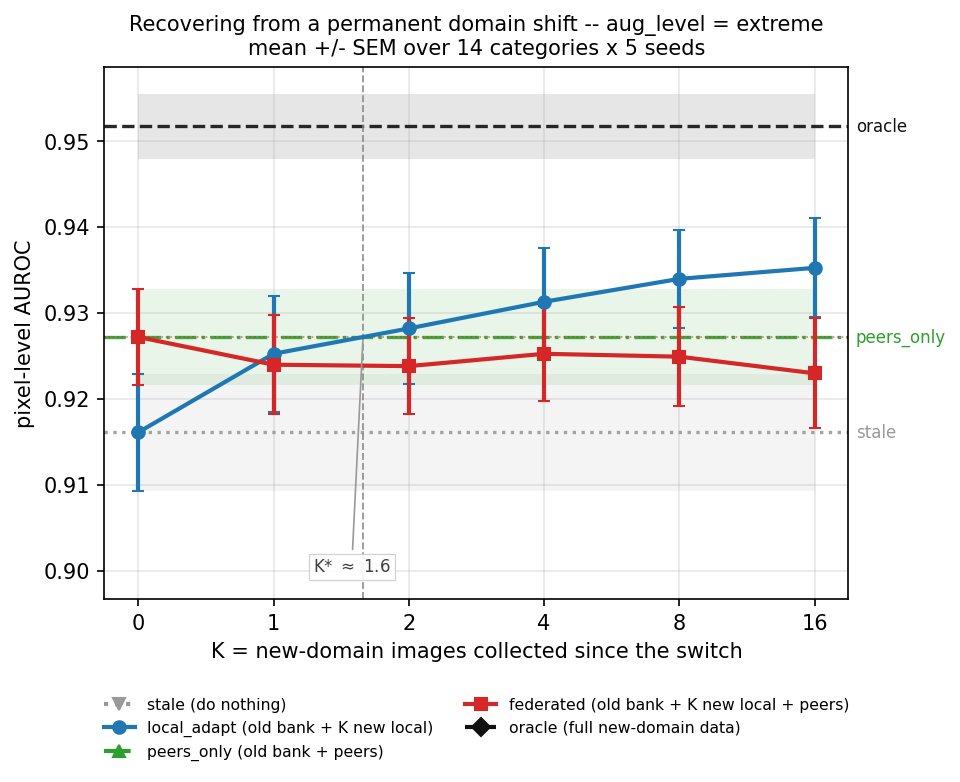}
    \caption{Pixel-level AUROC comparison after using extreme and severe non-IID data distributions}
    \label{fig:noniid}
\end{figure}

We analyze whether using non-IID data from other actors aids an actor in recovering performance with fewer new images than by just using their own data. To investigate this effect, a set of parametrizable photometric transforms (ColorJitter, brightness, contrast, saturation, hue, gaussian blur) is introduced, simulating differences in lighting and image capturing. The distribution shift and difference to other actors is achieved by assigning each actor its own randomly chosen transforms, while verifying that the mean embedding distance between two distributions is sufficiently large. The actor's performance is then evaluated on the domain-shifted full test set with the new transforms applied. Five different memory banks are considered: (1) the stale bank, (2) the stale bank + $K$ samples from the new distribution, (3) the stale bank + peers only, (4) the stale bank + $K$ samples from the new distribution + peers, and (5) an oracle built from the actor's full new-domain data. At image-level, using data from peers alone closes roughly half (extreme) to three quarters (severe) of the stale-to-oracle gap without the actor using any new samples. Federating on top of local adaptation improves performance most when the actor has little data and the benefit decreases as $K$ grows. At pixel-level, federating only benefits the actor at $K=0$, indicating that peer embeddings aid detection but dilute localization.

% \begin{figure}[!t]
%     \centering
%     \begin{minipage}{0.47\columnwidth}
%         \centering
%         \includegraphics[width=\linewidth]{images/auroc_pixel_severe.png}
%     \end{minipage}
%     \hspace{0.02\columnwidth}
%     \begin{minipage}{0.47\columnwidth}
%         \centering
%         \includegraphics[width=\linewidth]{images/auroc_pixel_extreme.png}
%     \end{minipage}
%     \caption{Pixel-level AUROC comparison after using extreme and severe non-iid data distributions}
%     \label{fig:noniid}
% \end{figure}

\textbf{Unreliable actor participation}
Actors might join the construction process at any point in time or be unavailable for certain periods. We investigate the implications on \textit{Federated Subsampling} under the unreliable availability of all actors in each round. We establish four scenarios: (baseline) - all actors are always available, unrealistic best-case;
(random) - a random feature across all actors is selected per round, simulating 
actors joining and leaving randomly; (skewed) - a $20\%$ probability of failure per actor is assumed in each round, resulting in a skewed distribution, depicting a more realistic scenario; (worst) - only the worst feature is selected. While the baseline scenario is not always yielding the best result due to the greedy nature of the approach, its performance stays consistent across the increase of actors in the system. Meanwhile, scenarios with more exploration are more unstable towards the number of actors in a system. Regardless of the strategy, the results indicate a more resilient behavior compared to other federated learning approaches.

\begin{table}[hb]
\centering
\caption{Mean AUROC $\pm$ stdev across all categories of the MVTec AD dataset for different participation settings}
\label{tab:client_results}
\setlength{\tabcolsep}{4pt}%
\begin{tabular}{l c c c}
\toprule
\textbf{Method} & {3 Actors} & {4 Actors} & {5 Actors} \\
\midrule
\textbf{Image-Level} & & & \\
\midrule
Baseline  & $97.46\pm 4.57$ & $97.46\pm4.57$ & $97.46\pm4.57$ \\
Random       & $97.93\pm 3.93$ & $97.83\pm3.97$ & $97.89\pm3.46$ \\
Skewed       & $97.71\pm 4.13$ & $97.95\pm3.40$ & $97.94\pm3.83$ \\
Worst & $97.41\pm 4.84$ & $97.01\pm5.79$ & $96.65\pm5.68$ \\
\midrule
\textbf{Pixel-Level}  & & \\
\midrule
Baseline  & $95.94\pm 2.94$ & $95.94\pm 2.94$ & $95.94\pm2.94$ \\
Random       & $96.06\pm 2.75$ & $96.05\pm 2.76$ & $95.99\pm2.81$ \\
Skewed       & $96.01\pm 2.85$ & $95.99\pm 2.91$ & $95.92\pm3.09$ \\
Worst & $96.17\pm 2.56$ & $96.24\pm 2.44$ & $96.02\pm2.78$ \\
\bottomrule
\end{tabular}
\end{table}

% \begin{table}[!t]
% \centering
% \caption{Average AUROC results $\pm$standard deviation across all categories of the MVTec AD dataset for different participation settings}
% \label{tab:client_results}
% \begin{tabular}{l c c c}
% \toprule
% \textbf{Method} & {3 Clients}  & {5 Clients} \\
% \midrule
% \textbf{Image-Level} &  & \\
% \midrule
% Always Best  & $97.46\pm 4.57$ & $97.46\pm4.57$ \\
% Random       & $97.93\pm 3.93$ & $97.89\pm3.46$ \\
% Skewed       & $97.71\pm 4.13$ & $97.94\pm3.83$ \\
% Always Worst & $97.41\pm 4.84$ & $96.65\pm5.68$ \\
% \midrule
% \textbf{Pixel-Level} & & & \\
% \midrule
% Always Best  & $95.94\pm 2.94$ & $95.94\pm2.94$ \\
% Random       & $96.06\pm 2.75$ & $95.99\pm2.81$ \\
% Skewed       & $96.01\pm 2.85$ & $95.92\pm3.09$ \\
% Always Worst & $96.17\pm 2.56$ & $96.02\pm2.78$ \\
% \bottomrule
% \end{tabular}
% \end{table}

\section{Conclusion}
We propose two novel approaches for construction of memory banks in federated settings. \textit{Double Subsampling} allows for a federated usage of memory banks, consolidating multiple localized memory banks, providing similar results compared to a fully local model by assuming the contents of a memory bank are similar to features extracted from a dataset. While this approach can be applied on top of existing methods, \textit{Federated Subsampling} introduces a strategy to federate the construction process of a memory bank model, allowing for a guaranteed consistent global model. Experiments show the resilience to non-IID distribution and improved cold-start performance.  Finally, the method proves resilient to non-deterministic actor behavior due to availability and timing.

\clearpage

\section{Acknowledgements}
This publication is co-financed by the European Union under the "IBW/EFRE \& JTF \"Osterreich 2021-2027" program and the State of Vorarlberg.

LLMs by Anthropic and Overleaf were used for grammar enhancement and minor rephrasing of sentences across the article. Additionally, coding agents by Anthropic were used for assistance in setting up experiments and supporting the creation of graphics. All code and text generated by an LLM was carefully revised and checked.

\bibliographystyle{IEEEbib}
\bibliography{FederatedSubsamplingPaperBibtex}

\end{document}